\documentclass[a4paper]{jpconf}
\usepackage{graphicx}
\usepackage{amsmath}

\begin{document}
\title{Geodesic Deviation Equation in Bianchi Cosmologies}

\author{D L C\'{a}ceres$^1$, L Casta\~{n}eda$^2$ and J M Tejeiro}

\address{Observatorio Astron\'{o}mico Nacional, Universidad Nacional de Colombia}

\ead{$^1$dlcaceresu@bt.unal.edu.co, $^2$lcastanedac@unal.edu.co}

\begin{abstract}
We present the Geodesic Deviation Equation (GDE) for the Friedmann-Robertson
Walker(FRW) universe and we compare it with the equation for
Bianchi type I model. We justify consider this cosmological model
due to the recent importance the Bianchi Models have as
alternative models in cosmology. The main property of these
models, solutions of Einstein Field Equations (EFE) is that they
are homogeneous as the FRW model but they are not isotropic. We can
see this because they have a non-null Weyl tensor, which is zero
for FRW model. We study some consequences of this Weyl tensor in
the GDE.
\end{abstract}

\section{Introduction}

Bianchi cosmological models are homogeneous solutions of EFE.
These models generalize FRW model, although they are
homogeneous, as FRW one, generally they are anisotropic. Standard
Cosmological Model has survived almost all observational tests
\cite{spergel}, however Bianchi cosmologies haven't been
discarded. Today the Universe is highly isotropic, but in early
times it couldn't been so, and it is an open question how
anisotropies died. Standard Model fails to explain quadrupole
anomaly \cite{rodrigues}, obtained from Cosmic Microwave
Background data. This anomaly have a plausible explanation with
Bianchi models, for that reason these models are considered today
\cite{rodrigues}. Thus we considered the idea of studying Geodesic
Deviation Equation(GDE) in some of these models.

\section{1+3 Covariant description}

We assume a fluid description for the cosmological model. The
4-velocity $u^{\alpha}$ of preferred worldlines is normalized,
$u_{\alpha}u^{\alpha}=-1$. Given this 4-velocity, we define the
projection tensor $h_{\alpha\beta}$, the expansion scalar
$\Theta$, the shear $\sigma_{\alpha\beta}$ and the vorticity
$\omega_{\alpha\beta}$ \cite{wainwright}, \cite{9812046}. In FRW
case $\sigma_{\alpha\beta}=\omega_{\alpha\beta}=0$. These
quantities are related with Weyl tensor. In FRW case Weyl tensor
is zero, so it is conformally flat, but it is not the case in
Bianchi models. In the $1+3$ covariant description we can split
the Weyl tensor into an electric $E_{\alpha\beta}$ and a magnetic
part $H_{\alpha\beta}$ \cite{9812046}, \cite{wainwright}.\\

\section{Deviation Equation}

In a space-time torsion free, if $V^{\alpha}$ is the normalized
tangent vector field of a fiducial geodesic, parameterized by an
affine parameter $\nu$, with $V_{\alpha}V^{\alpha}=\varepsilon$,
with $\varepsilon=+1,0,-1$ if the geodesics are spacelike, null or
timelike respectively; if $\eta^{\alpha}$ is the deviation vector
for the congruence, then the evolution of the separation or the
deviation vector $\eta^{\alpha}$ evolves according to GDE
\cite{9709060}:

\begin{equation}
\frac{\delta^{2}\eta^{\alpha}}{\delta\nu^{2}}=-R^{\alpha}_{\beta\gamma\delta}V^{\beta}\eta^{\gamma}V^{\delta}
\end{equation}

where $\frac{\delta T^{a...}_{b...}}{\delta
\nu}=V^{\gamma}\nabla_{\gamma}T^{a...}_{b...}$ is the covariant
derivative along the geodesic.\\

Now, let's see how is this equation in Bianchi type I model, given
by the metric\cite{0411080}, \cite{kumar}:

\begin{equation}
ds^{2}=dt^{2}-a^{2}(t)dx^{2}-b^{2}(t)dy^{2}-c^{2}(t)dz^{2}
\end{equation}

If we define $\tau(t):=a(t)b(t)c(t)$, from the energy-momentum
tensor conservation we get an analogous expression to a known
Friedmann equation \cite{0411080}, \cite{kumar}:

\begin{equation}
\frac{\ddot{\tau}}{\tau}=\frac{3\kappa}{2}(\mu-p)
\end{equation}

where $\mu$ is the Energy density and $p$ is the pressure. As we
see in the GDE, there is a difference if space-time has a nonzero
Weyl tensor. Ellis and van Elst in \cite{9709060} consider this
equation for FRW space-time.\\

For Bianchi type I model vorticity and Magnetic part of Weyl
tensor are equal to zero, $\omega_{\alpha\beta}=0$,
$H_{\alpha\beta}=0$. The Electric and shear tensors are non-null
and diagonal. For that reason, if we define
$F=E_{\beta\gamma}V^{\beta}\eta^{\gamma}$ and
$E=-V_{\alpha}u^{\alpha}$, if we suppose a perfect fluid,
$q^{\alpha}=0$, $\pi_{\alpha\beta}=0$ we have that GDE for this
space is:

\begin{equation}
\begin{split}
R^{\alpha}_{\beta\gamma\delta}V^{\beta}\eta^{\gamma}V^{\delta}=[\varepsilon\frac{1}{3}(\mu+\Lambda)+\frac{1}{2}(\mu+p)E^{2}]\eta^{\alpha}\\
+F(2u^{\alpha}+\eta^{\alpha}-V^{\alpha})+E^{\alpha}_{\gamma}\eta^{\gamma}(\varepsilon+2E^{2})
\end{split}
\end{equation}

where $\Lambda$ is the cosmological constant. The first part of
this expression is the force term representing the perfect part of
the fluid \cite{9812046} and only gives isotropic deviation. In
FRW case we only have this term and reflects spatial isotropy of
space-time, deviation vector only changes in magnitude but not in
direction. The other terms change the direction of
$\eta^{\alpha}$.\\

If we suppose an equation of state $p=w\epsilon$, ($w=0$ for dust,
$w=\frac{1}{3}$ for radiation), then, for $\Lambda=0$ we
get\cite{0411080}:

\begin{equation}
\tau=At^{\frac{2}{1+w}}
\end{equation}

Given $\tau$ as a function of t we can get $a_(t)$, $b(t)$ and
$c(t)$. The electric Weyl tensor tends to zero. It is diagonal and
its components decay. For this model the vorticity and the
Magnetic part of Weyl tensor are null. The EFE imply that
$q^{\alpha}=0$. This means we have a fluid without viscosity.
However, we can have an imperfect fluid because the model allows
to have an anisotropic pressure $\pi_{\alpha\beta}$. But this
anisotropic pressure tensor must be diagonal, following EFE's.
Given $\tau$ as a function of $t$ we have expressions for $a(t)$,
$b(t)$ and $c(t)$. The evolution of the three scale factors for
$w=0$ is shown in Figure 1 and for $w=\frac{1}{3}$ in figure 2. We
can see that in radiation case anisotropy is greater than in
matter dominated case, but in both cases anisotropy dies.\\

The $\sigma$ scalar defined by
$\sigma^{2}=\frac{1}{2}\sigma^{\alpha\beta}\sigma_{\alpha\beta}$
also measure the degree of anisotropy in the model. In Figure 4
we illustrate the evolution of $\sigma^{2}$.\\

\section{Conclusions}

The evolution of the Electric Weyl tensor show us that this model
tends to isotropy as t grows, the model isotropizes. Also the
evolution of shear scalar. In FRW only the magnitude $\eta$ will
change along a geodesic, while its spatial orientation remain
fixed \cite{9709060}. So, in early times the deviation is
anisotropic, the orientation is important, for example, along the
z axis deviation will grow, while along x and y axis the effect of
Weyl Tensor is to contract the deviation, because there are the
other terms that are also present in FRW. Here we considered a
perfect fluid, $\pi_{\alpha\beta}=0$. Along x axis, the effect of
anisotropy is minimal.

\begin{figure}
\begin{minipage}{1.7in}
\includegraphics[width=1.7in]{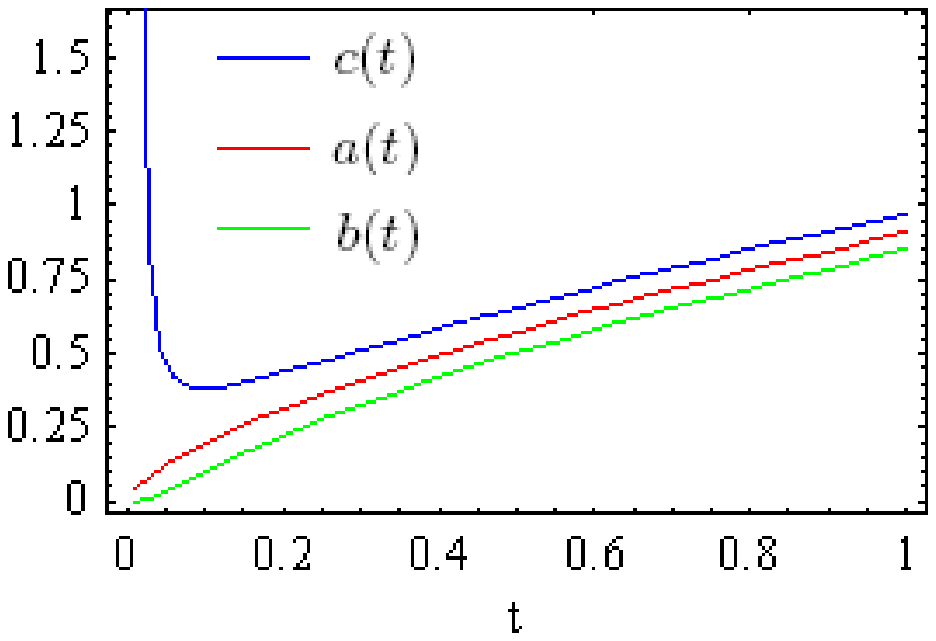}
\caption{\label{label}Scale factors for $w=0$, $A=\frac{3}{4}$.}
\end{minipage}\hspace{.5in}
\begin{minipage}{1.7in}
\includegraphics[width=1.7in]{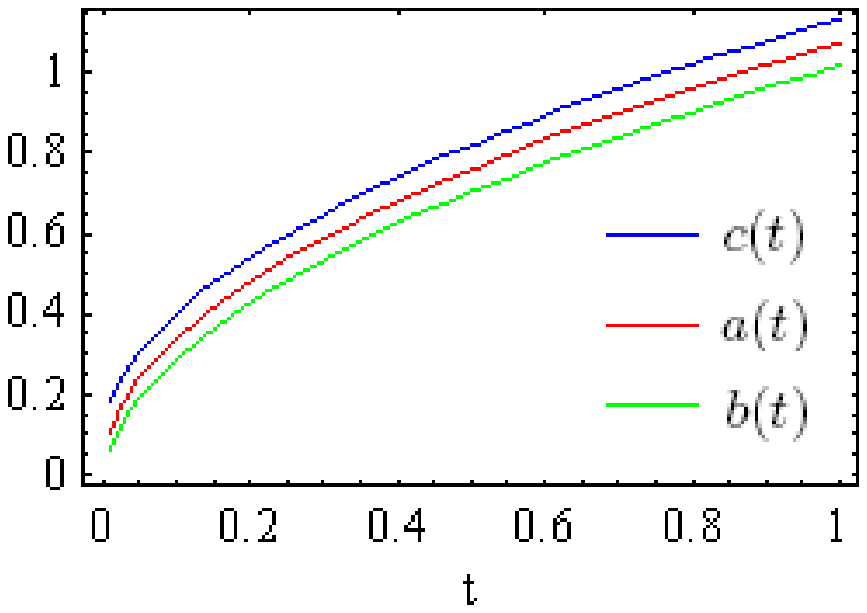}
\caption{\label{label}Scale factors for $w=\frac{1}{3}$,
$A=\frac{3}{4}$.}
\end{minipage}\hspace{.5in}
\begin{minipage}{1.7in}
\includegraphics[width=1.7in]{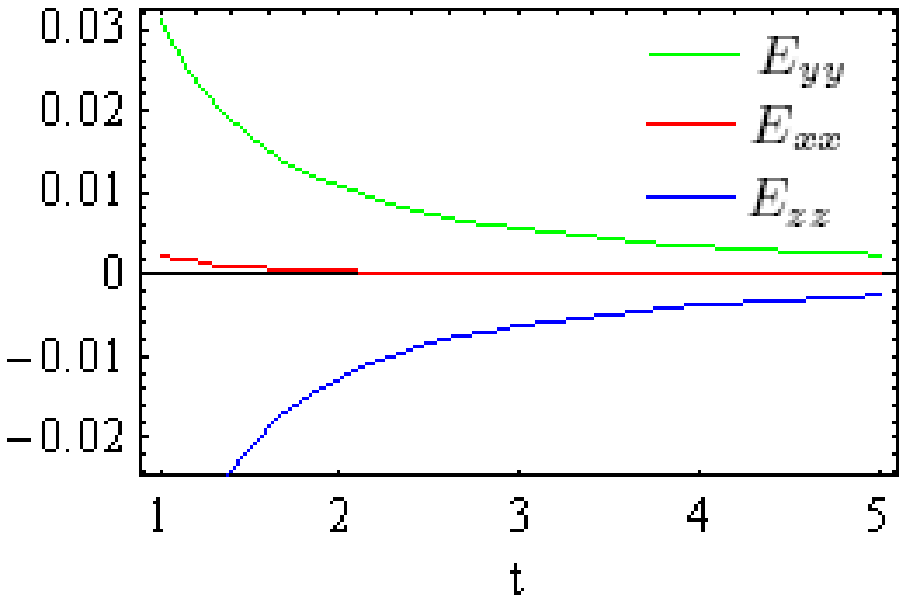}
\caption{\label{label}Electric Weyl tensor for $A=\frac{3}{4}$,
$w=\frac{1}{3}$.}
\end{minipage}
\end{figure}

\begin{figure}[h]
\begin{minipage}{2in}
\includegraphics[width=2in]{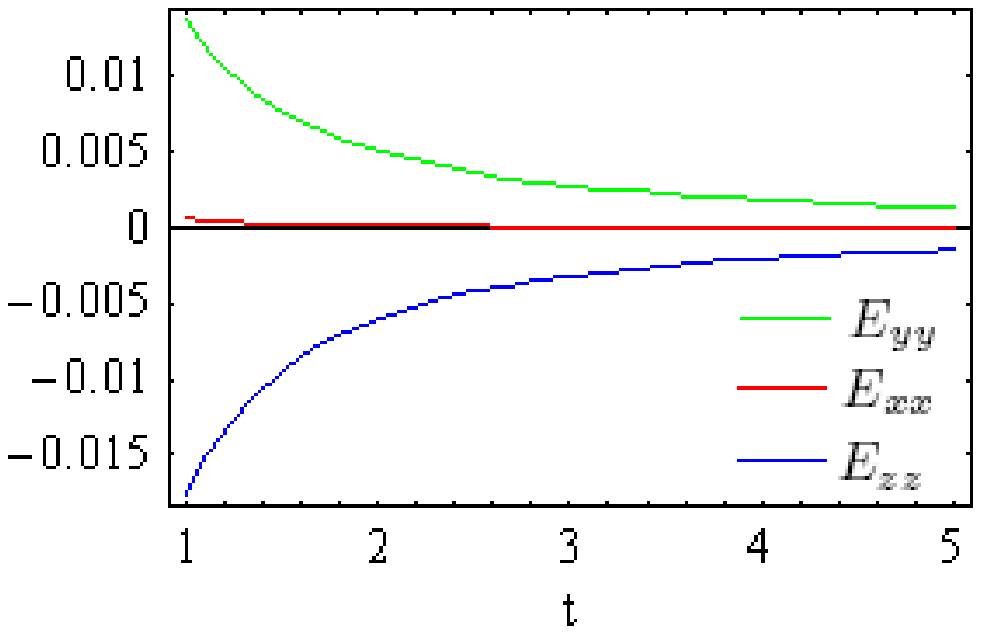}
\caption{\label{label}Electric Weyl tensor for $A=\frac{3}{4}$,
$w=\frac{1}{3}$.}
\end{minipage}\hspace{1in}
\begin{minipage}{2in}
\includegraphics[width=2in]{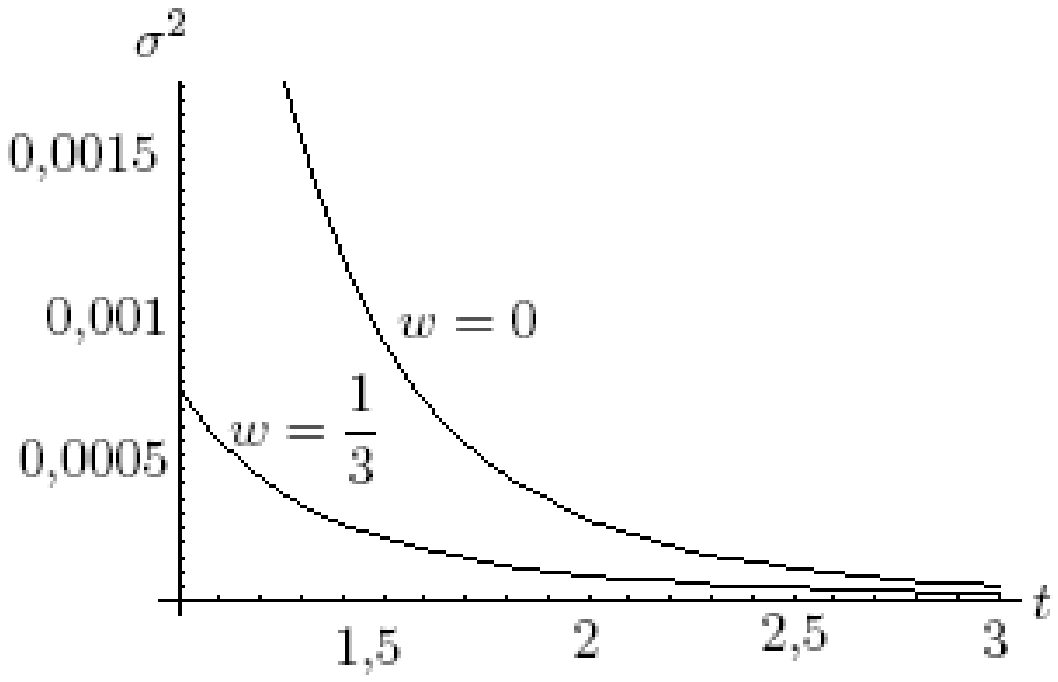}
\caption{\label{label}Shear scalar evolution for $A=\frac{3}{4}$.}
\end{minipage}
\end{figure}


\end{document}